\documentclass[12pt]{article}
\usepackage{graphicx}
\usepackage{bm}
\usepackage[utf8]{inputenc}
\usepackage{lmodern}  
\begin{document}
\title{Time-Dependent Gaussian Solution for the Kostin
Equation around Classical Trajectories} 
\date{}
\maketitle 
\begin{center}
\large{F. Haas 
\vskip.3cm
Departamento de F{\'i}sica, Universidade Federal do Paran\'a, 81531-990, Curitiba, Paran\'a, Brazil

\vskip.5cm
J. M. F. Bassalo
\vskip.3cm
Funda\c{c}\~ao Minerva,\ Av. Governador Jos\'e Malcher\ 629,  
             66035-100,\ Bel\'em,\ Par\'a,\ Brazil
             
\vskip.5cm
D. G. da Silva
\vskip.3cm
Escola Munguba do Jari, Vit\'oria do Jari, 68924-000, Amap\'a,\ Brazil             

\vskip.5cm
A. B. Nassar
\vskip.3cm
Extension Program - Department of Sciences, University of
California, Los Angeles, California 90024,\ USA          

\vskip.5cm
M. Cattani
\vskip.3cm
Universidade de S\~ao Paulo, Instituto de F\'{\i}sica, 05315-970,\ S\~ao Paulo,\ SP, Brazil}

   \end{center}
   
   \newpage
\begin{abstract}{The structure of time-dependent Gaussian solutions for the Kostin equation in dissipative quantum mechanics is analyzed. Expanding the generic external potential near the center of mass of the wave packet, one conclude that: the center of mass follows the dynamics of a classical particle under the external potential and a damping proportional to the velocity; the width of the wave packet satisfy a non-conservative Pinney equation. An appropriate perturbation theory is developed for the free particle case, solving the long standing problem of finding analytic expressions for square integrable solutions of the free Kostin equation. The associated Wigner function is also studied. }
\end{abstract}
\vskip.3cm
%






\maketitle

\section{Introduction}
\label{intro}
Presently there is no universal consensus on how to quantize dissipative classical systems, a subject of great practical and fundamental interest. Among the available alternatives, the Kostin logarithmic nonlinear Schr\"odinger equation \cite{Kostin} provides a suitable modeling for dissipation in quantum mechanics. We note that presently there are no experiments measuring deviations,
if any, from linearity in quantum theory. Hence, in spite of its drawbacks (breaking of the superposition principle, wrong modified frequency in the case of damped harmonic oscillations \cite{Hasse}), it is of interest to have a deeper understanding of the solutions of the Kostin equation. Moreover, recently the Kostin equation (also termed Schr\"odinger-Langevin equation) is attracting attention, in the context of dissipative time-dependent density functional theory \cite{Tsekov2}. Similarly, the Kostin logarithmic term was suggested as an appropriate bath functional in time-dependent density functional theory for open quantum systems with unitary propagation \cite{Yuen}. In addition, some recent studies include a rigorous mathematical analysis of the Kostin equation in bounded domains \cite{Lopez}, the stability analysis of the equilibrium states \cite{Van}, the search for numerical solutions under diverse confinement potentials \cite{Sanin}, as well as connections on quantum combinatorial optimization, or quantum annealing \cite{Falco}. The existence of soliton-like solutions for Schr\"odinger-Langevin equations in a fluid form from the perspective of Nelson's stochastic mechanics \cite{Nelson} has been also investigated \cite{Nassar}. 

The interest on the Kostin model arises because: it preserves the norm of the wave function, in spite of the intrinsic nonlinearity; when written in the hydrodynamic form, one find a linear in the fluid velocity friction term, with a simple interpretation in terms of Ohmic dissipation; the Ehrenfest relations for dissipative systems are strictly satisfied. Further possible difficulties, relying on the existence of undamped stationary states, actually can be understood since these states might be interpreted as one-dimensional projections
of the normal modes of motion of a system with many degrees of freedom coupled by conservative forces \cite{Immele}.

It is a long standing problem, to accurately describe the free particle dynamics in the Kostin model and so far only numerical solutions are available \cite{Hasse}, \cite{Immele}. It is one of the main purposes of this work, to circumvent this difficulty by means of an appropriate perturbation theory in a damping dominated scenario. Furthermore, it was shown \cite{Brull} that in the damped free particle case the Kostin equation can have no solitary wave type solutions of the form $\Psi = \Psi(x+ct)$  case which lie in $L^2$, where $c$ is a constant. Also, in his original paper Kostin \cite{Kostin} have constructed some solutions of the plane wave like form $\Psi = \exp(i\theta(t))\exp[(i/\hbar)(\alpha(t)x+\beta(t))]$, which are however also not square integrable. Here by ``damped free particle" case it is mean the dissipative motion under a null external potential. 

This paper is organized as follows. In Section II, we present the Kostin model in hydrodynamic form. In Section III, a time-dependent Gaussian {\it An\-satz} for the solution of the Kostin equation is proposed. After expanding the arbitrary external potential near the center of mass of the wave packet, it is found that the center of mass solves the same Newtonian equation as the one for a particle under the external potential and a damping force proportional to the velocity. In addition, the width of the wave packet is shown to solve the damped Pinney equation \cite{Haas}, \cite{Pinney}, a nonlinear equation whose general properties are not fully understood until now. In Section IV, the free particle case is analyzed in detail by means of perturbation theory. In Section V the obtained solutions are discussed in terms of the Wigner function formalism. Section VI is dedicated to the conclusions. 

\section{Kostin model in hydrodynamic form}
\label{sec:1}
In 1972, Kostin \cite{Kostin} proposed the following logarithmic nonlinear Schr\"{o}dinger
equation to represent dissipative physical systems,
\begin{eqnarray}
\label{e1}
i\hbar\frac{\partial\Psi}{\partial t} &=& - \frac{\hbar^2}{2m}\frac{\partial^2\Psi}{\partial x^2} + \left[V - i\frac{\hbar\nu}{2}\ln\Big{(}\frac{\Psi}{\Psi^{*}}\Big{)}
+ i\frac{\hbar\nu}{2}\Big{<}\ln\Big{(}\frac{\Psi}{\Psi^{*}}\Big{)}\Big{>}\right]\Psi \,, 
\end{eqnarray}
where $\Psi = \Psi(x,t)$ and $V = V(x,t)$ are, respectively, the normalized 
wave function and the external potential, $\hbar$ is Planck's constant divided by $2\pi$, $m$ is the mass and ${\nu}$ is a positive damping parameter corresponding to Ohmic dissipation. For simplicity, in this work only one-dimensional problems are considered. Moreover, in Eq. (\ref{e1}) we have 
\begin{equation}
\label{e2}
\Big{<}\ln\left(\frac{\Psi}{\Psi^{*}}\right)\Big{>} = \int dx |\Psi(x,t)|^2 \ln\left(\frac{\Psi(x,t)}{\Psi^{*}(x,t)}\right) \,,
\end{equation}
a term introduced in such a manner so that the expectation value for
the energy becomes just the sum of the kinetic and potential parts. A potential term associated to a random force due to 
interaction with a heat bath could have been also added, as in the original Kostin formulation \cite{Kostin}; for simplicity, here this possibility is disregarded.

We introduce the Madelung \cite{Madelung} decomposition
\begin{equation}
\label{e3}
\Psi = A\exp(iS/\hbar) \,,
\end{equation}
where the amplitude $A = A(x,t)$ and the phase $S = S(x,t)$ are real functions, and also the probability density $\rho$ and the quantum fluid velocity $v$ according to 
\begin{equation}
\label{e4}
\rho = A^2 \,, \quad v = \frac{1}{m}\frac{\partial S}{\partial x} \,.
\end{equation}
From this prescription and separating the real and imaginary parts of the Kostin equation (\ref{e1}) gives 
\begin{eqnarray}
\label{e5}
\frac{\partial\rho}{\partial t} + \frac{\partial}{\partial x}(\rho v) &=& 0 \,,\\
\label{e6}
\frac{\partial v}{\partial t} + v\frac{\partial v}{\partial x} &=& - \frac{1}{m}\frac{\partial V}{\partial x} - \nu v + \frac{\hbar^2}{2m^2}\frac{\partial}{\partial x}\left(\frac{\partial^{2}\sqrt{\rho}/\partial x^2}{\sqrt{\rho}}\right) \,,
\end{eqnarray}
which are formally the same as resp. the continuity and force equations for a fluid acted by an external potential V, also including a damping term proportional to the velocity and a quantum Bohm potential term (the $\sim \hbar^2$ contribution in Eq. (\ref{e6})). 

In terms of the hydrodynamic formulation, the existence of undamped solutions $\Psi = \varphi_{n}(x)\exp(-i E_{n}t/\hbar)$, where $\varphi_{n}(x)$ and $E_n$ are resp. the stationary states and eigenvalues of the Hamiltonian operator $H = - (\hbar^{2}/[2m])\partial^{2}/\partial x^2 + V(x)$ in the autonomous $V = V(x)$ case, can be understood since these solutions have zero quantum fluid velocity and hence suffer no friction.

\section{Expansion around classical trajectories}
\label{sec:2}
We propose the following Gaussian {\it Ansatz} for the probability density, 
\begin{equation}
\label{e7}
\rho = \frac{1}{\sqrt{2\pi}\,a}\exp\left(- \frac{(x-q)^2}{2a^2}\right) \,,
\end{equation}
where $a = a(t)$ and $q = q(t)$ are auxiliary functions of time to be determined in what follows, interpreted resp. as the width and the center of mass of the wave packet. Using Eq. (\ref{e7}), the only velocity field satisfying the continuity equation with a vanishing probability current density $\rho v$ at infinity turns out to be given by 
\begin{equation}
\label{e8}
v = \frac{\dot{a}}{a}(x-q) + \dot{q} \,.
\end{equation}

Gaussian wave packet solutions to the Kostin equation have been discussed by many authors, specially in the harmonic confinement case \cite{Brull}, \cite{Hasse}, \cite{Immele}, \cite{Manko}. In this work we highlight the generic applicability of this approach, in the following sense. Consider the Taylor expansion 
\begin{equation}
\label{e9}
V(x,t) = V(q(t),t) + V'(q(t),t) (x-q(t)) + \frac{V''(q(t),t)}{2} (x-q(t))^2 + \dots \,,
\end{equation}
where a prime denotes derivative with respect to the first argument. Retaining up to quadratic terms in $V$, substituting in the force equation (\ref{e6}) and separating terms proportional to $(x-q)^0$ and $(x-q)$ gives
\begin{eqnarray}
\label{e10}
\ddot{q} + \nu\dot{q} &=& - \frac{1}{m}V'(q(t),t) \,,\\
\label{e11}
\ddot{a} + \nu\dot{a} &+& \omega^{2}(t)a = \frac{\hbar^2}{4m^2 a^3} \,,
\end{eqnarray}
where we have defined $\omega^{2}(t) = V''(q(t),t)/m$. Hence the center of mass $q$ can be identified with the classical particle under the potential $V$ and a linear in velocity friction. It should be stressed that Eq. (\ref{e10}) is just the Ehrenfest theorem for the
potential (\ref{e9}) and distribution ({\ref{e7}). On the other hand the width $a$ satisfy the damped Pinney equation (\ref{e11}). In the conservative ($\nu = 0$) case the (nonlinear) Pinney equation is well known to be exactly solvable in terms of the solution of the (linear) time-dependent harmonic oscillator equation \cite{Pinney}. Approximate solutions exist \cite{Haas} for weak damping and slowly varying non-vanishing frequencies $\omega(t)$. 

The eikonal $S$ can be found integrating Eq. (\ref{e4}) with velocity field given by Eq. (\ref{e8}), which then gives the wave function
\begin{equation}
\label{e12}
\Psi = \frac{1}{(2\pi a^2)^{1/4}}\exp\left[-\frac{(x-q)^2}{4a^2}\right]\exp\left[\frac{im}{\hbar}\left(\frac{\dot{a}}{2a}(x-q)^2 + \dot{q}(x-q)\right)\right] \,.
\end{equation}
An extra arbitrary function of time could have been added to $S$, but we disregard it since it does not change neither the probability density nor the pro\-ba\-bi\-li\-ty current. 

The generic character of the approximate solution (\ref{e12}) results from the arbitrariness of the external potential, as long as the expansion near classical trajectories (\ref{e9}) holds. The remaining task is to evaluate $q, a$ from Eqs. (\ref{e10}) and (\ref{e11}), which can be done only in special cases. Curiously, even the free particle ($V = 0$) case does not seems to be solvable in closed form. In the next Section a perturbative solution is build for this situation. 

\section{The free particle case}
\label{sec:4}
Assuming $V = 0$, Eq. (\ref{e10}) is trivially solved as
\begin{equation}
\label{e13}
q = q_0 + \frac{\dot{q}_0}{\nu}\,[1-\exp(-\nu t)] \,,
\end{equation}
where $q(0) = q_0$, $\dot{q}(0) = \dot{q}_0$. On the other hand, the solution for the width equation
\begin{equation}
\label{e14}
\ddot{a} + \nu\dot{a} = \frac{\hbar^2}{4m^2 a^3} 
\end{equation}
apparently can not be analytically found. Moreover it can be checked that the simple approximate expressions from Ref. \cite{Haas} developed for a non-zero harmonic confinement did not apply in the free case. 

The ultimate reason why Eq. (\ref{e14}) is hard to solve, even approximately for small damping, is because the unperturbed solution (for $\nu = 0$) grows without bound as $t \rightarrow \infty$. Indeed, in this conservative case the general solution is
\begin{equation}
\label{fr}
a^2 = \left(a_0 + \dot{a}_{0}t\right)^2 + \frac{\hbar^2 t^2}{4m^2a_{0}^2} \,, \quad (\nu = 0) 
\end{equation}
where $a(0) = a_0, \dot{a}(0) = \dot{a}_0$. 

On the other hand, in the damping dominated case where we can neglect the acceleration in Eq. (\ref{e14}) and integrating the resulting first-order ordinary differential equation, we get 
\begin{equation}
\label{asy}
a \sim \left(\frac{\hbar^2 t}{m^2 \nu}\right)^{1/4} \,,
\end{equation}
for large times, which is consistent with results found from nonlinear theories of quantum Brownian motion \cite{Tseko}, \cite{Tsekov}.
As pointed out in Ref. \cite{Razavy} in the analysis of wave packet spreading in a viscous medium, the leading edge experiences a larger friction force than the interior of the Gaussian. In this way a retardation of the spreading takes place, which explains the different power laws $a \sim t$ for the non-dissipative case in Eq. (\ref{fr}) and $a \sim t^{1/4}$ for the dissipative case in Eq. (\ref{asy}). Here it is proposed a different framework for this point, in terms of the collision-dominated regime of the damped Pinney equation without harmonic term.

On general grounds, without neglecting the acceleration term, we have 
\begin{equation}
\frac{d}{dt}\left(\frac{\dot{a}^2}{2} + \frac{\hbar^2}{8m^2 a^2}\right) = - 2\nu\dot{a}^2 \leq 0 \,.
\end{equation}
Since every term inside brackets is positive definite, the kinetic energy term as well as the inverse square term tend to be negligible for large times. This is consistent with the asymptotic result (\ref{asy}). Moreover, from Eq. (\ref{asy}) we find
\begin{equation}
\frac{\ddot{a}}{\nu\dot{a}} \sim - \frac{3}{4\nu t} \,,
\end{equation}
further justifying the neglect of the acceleration term as long as $\nu t \gg 1$.

To start developing a perturbation solution, first rescale according to 
\begin{equation}
\label{e15} 
t \rightarrow \nu t \,, \quad a \rightarrow \left(\frac{m\nu}{\hbar}\right)^{1/2} a \,,
\end{equation}
so that Eq. (\ref{e14}) is rewritten as
\begin{equation}
\label{e16}
\epsilon\ddot{a} + \dot{a} = \frac{1}{4a^3} \,,
\end{equation}
where $\epsilon$ is a perturbation parameter eventually set to unity at the end of the calculation. The same symbols are used for the original and dimensionless quantities, for simplicity of notation. The parameter $\epsilon$ is introduced in Eq. (\ref{e16}), in accordance with our previous considerations indicating that the acceleration term can be viewed as a perturbation, except for a transient time.

We then define
\begin{equation}
a = \alpha_0 + \epsilon \alpha_1 + O(\epsilon^2) \,,
\end{equation}
where $\alpha_0, \alpha_1$ are functions to be determined. Inserting in Eq. (\ref{e16}) and equating terms of equal power of $\epsilon$ we get
\begin{eqnarray}
\label{e17}
\dot{\alpha}_0 &=& \frac{1}{4\alpha_{0}^3} \,,\\
\label{e18}
\dot{\alpha}_1 + \frac{3 \alpha_1}{4\alpha_{0}^4} &=& - \ddot{\alpha}_0 \,.
\end{eqnarray}
For simplicity we restrict to first order. Higher order corrections are easily found, if necessary.

Equation (\ref{e17}) is a first order separable ordinary equation, with general solution
\begin{equation}
\label{e19}
\alpha_0 = (t + c_{1}^4)^{1/4} \,,
\end{equation}
where $c_1$ is the integration constant, left indeterminate for the moment. 

Given $\alpha_0$, Eq. (\ref{e18}) is a first order linear inhomogeneous equation for $\alpha_1$. Using Eq. (\ref{e19}), the solution is 
\begin{equation}
\label{e20}
\alpha_1 = \frac{c_2 + (3/16)\ln(t + c_{1}^4)}{(t+c_{1}^4)^{3/4}} \,,
\end{equation}
where $c_2$ is a further integration constant. 

In conclusion, setting $\epsilon = 1$ we find the approximate general solution
\begin{equation}
\label{e21}
a = \alpha_0 + \alpha_1 = (t + c_{1}^4)^{1/4} + \frac{c_2 + (3/16)\ln(t + c_{1}^4)}{(t+c_{1}^4)^{3/4}} \,.
\end{equation}
The two numerical constants $c_1$ and $c_2$ are determined from the initial conditions, 
\begin{equation}
\label{e22}
a(0) = c_1 + \frac{c_2}{c_{1}^3} + \frac{3\ln c_1}{4c_{1}^3} \,, \quad \dot{a}(0) = \frac{3-12c_2 + 4c_{1}^4 - 9\ln c_1}{16c_{1}^7} \,,
\end{equation}
where we assumed $c_{1}>0$. It turns out that the system (\ref{e22}) can be numerically solved only.

Equation (\ref{e21}) confirm the asymptotic behavior $a \sim t^{1/4}$ for large $t$. Further corrections can be constructed extending the method to second order but the result is increasingly awkward without significant gain. 

Eliminating $c_2$ from Eq. (\ref{e22}) results in 
\begin{equation}
\dot{a}(0) = \frac{1}{c_{1}^3} - \frac{3a(0)}{4c_{1}^4} + \frac{3}{16c_{1}^7} \,,
\end{equation}
which may admit several real solutions after specifying $a(0), \dot{a}(0)$. For instance, when $a(0) = 2, \dot{a}(0) = 0$ we get $c_1 = 1.437$ and then $c_2 = 1.399$. The numerical solution of Eq. (\ref{e16}) for these values can be checked to be virtually indistinguishable from the perturbation solution (\ref{e21}). However, if the second admissible set $c_1 = 0.591, c_2 = 0.685$ is used, the first-order perturbation solution gives not the same excellent agreement with the numerical solution, although reproducing the asymptotic behavior. 

Notice that by construction our approximate solution is adapted to the collision dominated regime, with the acceleration term treated as a perturbation. Hence for certain initial conditions one may have some transient disagreement between the proposed form and the numerical or the actual exact solutions. This is specially true for an initially rapidly expanding wave packet, where the effects of the damping become dominant only after a relatively long time scale. These considerations are verified through the direct numerical solution of the damped Pinney equation, with or without harmonic confinement term \cite{Tsekov3}.

\section{Wigner function}
It is interesting to briefly consider the Wigner function \cite{Wigner}, which is a valuable tool for the direct interpretation of the quantum system, on phase space. For a pure state the Wigner function $f = f(x,p,t)$ is defined as
\begin{equation}
\label{e23}
f = \frac{1}{2\pi\hbar}\int\,dy\,\Psi^{*}\left(x+\frac{y}{2},t\right)\,\exp\left(\frac{ip \,y}{\hbar}\right)\,\Psi\left(x-\frac{y}{2},t\right) \,,
\end{equation}
where $p$ is the momentum variable. From Eq. (\ref{e23}) we have $\int dx dp f(x,p,t) = 1$. Using the Gaussian form in Eq. (\ref{e12}) we derive
\begin{equation}
\label{e24}
f = \frac{1}{\pi\hbar}\exp\left[- \frac{(x-q)^2}{2a^2} - \frac{2a^2}{\hbar^2}\left(p-m\dot{q}-\frac{m\dot{a}}{a}(x-q)\right)^2\right] \,,
\end{equation}
which is valid near classical trajectories for any external potential as long as the pair $q, a$ satisfy Eqs. (\ref{e10}) and (\ref{e11}). Notice that this Wigner function is positive definite and has a Gaussian shape as well.

From Eq. (\ref{e24}) it is tempting to adopt the canonical transformation
\begin{equation}
x \rightarrow x - q \,, \quad p \rightarrow p - m\dot{q} - \frac{m\dot{a}}{a}(x-q) \,,
\end{equation}
so that in the transformed variables the level curves of the Wigner function become ellipses with semi-axis proportional to $a$ and $\hbar/a$. Hence the associated area is constant and proportional to $\hbar$, irrespective of the time evolution of the width $a$. However, clearly these results are a result from the specific class of the wave functions, independently of the dynamics generated by the external potential.

In passing, we note that the standard definitions $(\Delta x)^2 = <x^2> - <x>^2$ and $(\Delta p)^2 = <p^2> - <p>^2$ can be applied, with averages defined as in Eq. (\ref{e2}). Restoring physical coordinates the result is
\begin{equation}
\label{e25}
(\Delta x)^2 = a^2 \,, \quad (\Delta p)^2 = m^2 \dot{a}^2 + \frac{\hbar^2}{4a^2} \,,
\end{equation}
which is in accordance with the uncertainty principle since $\Delta x \Delta p \geq \hbar/2$. This is an advantage over other approaches using a linear, time-dependent Hamiltonian to describe dissipation, like the Kanai-Caldirola model \cite{Caldirola}, \cite{Kanai}, which is well-known to violate the uncertainty principle \cite{Brittin}.

\section{Conclusion}
Assuming a Gaussian solution for the Kostin equation and expanding near classical trajectories, in this work we have shown that the center of mass of the wave packet follows the classical damped dynamics under the external potential. Similarly, the width of the Gaussian is shown to satisfy Eq. (\ref{e11}), which is a non-conservative Pinney equation. Although closed form solutions can perhaps exist for special classes of external potentials, there are well-known obstacles to the integrability of the damped Pinney equation: it does not match the Tresse-Cartan conditions and hence is not linearizable under general point transformations; it does not possess quadratic in velocity cons\-tants of motion; it can be reduced to an Emden-Fowler equation of index $-3$, but with time-dependent coefficients so that it does not match known solvable cases; the use of generalized Sundman transformations may also be proven to be useless. See \cite{Haas} for a more detailed account on the non-solvability of Eq. (\ref{e11}) using diverse techniques. Hence in Section IV a perturbation approach in the friction dominated case is used, in order to derive for the first time analytical approximate solutions of the Kostin equation under zero external potential. We note that the same approximate expressions apply in the case of a purely time-dependent external force $F(t)$, because in this situation one has $V(q(t),t) = - F(t)q(t)$, $\omega(t) = 0$ so that Eq. (\ref{e14}) still holds. To conclude, the general structure of the solutions of Eqs. (\ref{e10}) and (\ref{e11}), which are ge\-ne\-ri\-cal\-ly representative in the Kostin model near classical trajectories, remains presently largely unexplored.

\vskip.9cm
{\bf Acknowledgments}: this work was supported by Conselho Nacional de Desenvolvimento Cient\'{\i}fico e Tecnol\'ogico (CNPq).

\end{document}